\documentclass{emulateapj} 
\usepackage{graphicx}
\usepackage{amsmath} 
\usepackage{apjfonts} 
\usepackage{amssymb} 
\usepackage{xspace}
\usepackage{hyperref} 
\usepackage{natbib} 
\usepackage{multirow}
\usepackage{epstopdf}
\usepackage{epsf}
\usepackage{subfigure}
\usepackage{epsf}
\usepackage{epstopdf}
\usepackage[usenames,dvipsnames]{xcolor}

\def\Mpc{{\rm Mpc}}
\def\kpc{{\rm kpc}}
\begin{document} 
\shorttitle{Non-Equilibrium Electrons in Cluster Outskirts} 
\shortauthors{Avestruz et al.}
\submitted{The Astrophysical Journal} 
\slugcomment{The Astrophysical Journal, accepted}

\title{Non-Equilibrium Electrons in the Outskirts of Galaxy Clusters}

\author{ Camille Avestruz\altaffilmark{1,2} }
\author{ Daisuke Nagai\altaffilmark{1,2,3} }
\author{ Erwin T. Lau\altaffilmark{1,2} }
\author{ Kaylea Nelson\altaffilmark{2,3}}
\affil{
$^1${Department of Physics, Yale University, New Haven, CT
  06520, U.S.A.;
  \href{mailto:camille.avestruz@yale.edu}{camille.avestruz@yale.edu}}\\
$^2${Yale Center for Astronomy \& Astrophysics, Yale
  University, New Haven, CT 06520, U.S.A.} \\
$^3${Department of Astronomy, Yale University, New Haven,
  CT 06520, U.S.A.}  \\
}

  
%-------------------------------------------------%
\begin{abstract} 
%-------------------------------------------------%
The analysis of X-ray and Sunyaev-Zel'dovich measurements of the
intracluster medium (ICM) assumes that electrons are in thermal
equilibrium with ions in the plasma. However, in the outskirts of
galaxy clusters, the electron-ion equilibration timescale can become
comparable to the Hubble time, leading to systematic biases in cluster
mass estimates and mass-observable scaling relations. To quantify an
upper limit of the impact of non-equilibrium electrons, we use a
mass-limited sample of simulated galaxy clusters taken from a
cosmological simulation with a two-temperature model that assumes the
Spitzer equilibration time for the electrons and ions. We show that
the temperature bias is more pronounced in more massive and rapidly
accreting clusters. For the most extreme case, we find that the bias
is of order $10\%$ at half of the cluster virial radius and increases
to $40\%$ at the edge of the cluster. Gas in filaments is less
susceptible to the non-equilibrium effect, leading to azimuthal
variations in the temperature bias at large cluster-centric
radii. Using mock {\em Chandra} observations of simulated clusters, we
show that the bias manifests in ultra-deep X-ray observations of
cluster outskirts and quantify the resulting biases in hydrostatic
mass and cluster temperature derived from these observations. We
provide a mass-dependent fitting function for the temperature bias
profile, which can be useful for modeling the effect of electron-ion
equilibration in galaxy clusters.
  \end{abstract}

%-------------------------------------------------%
\section{Introduction}
%-------------------------------------------------%

Galaxy clusters constrain cosmology, but the power of cluster-based cosmology 
is limited by our understanding of astrophysical processes that govern cluster formation.
The evolution of cluster abundance is particularly sensitive to cosmological parameters 
\citep[e.g.][for review]{allen_etal11}. Recent advances in X-ray and microwave 
observations have enabled measurements of properties of the intracluster medium 
(ICM) out to the virial radius and produced cosmological constraints that are 
complementary to other probes \citep{vikhlinin_etal09,mantz_etal10,planck_XX14}.

However, recent X-ray and microwave measurements at the largest radii have
exhibited unexpected features, which raise a question whether we understand the 
global ICM properties well enough to use galaxy clusters for a robust cosmological 
probe. First, several entropy profiles measured with the {\em Suzaku} X-ray observatory 
showed flattening outside of ${R}_{500c}$\footnote{$R_{500c}$ is the cluster radius 
enclosing an average density $500$ times the critical density of the universe.}
\citep[e.g.,][]{bautz_etal09,reiprich_etal09,
  hoshino_etal10,kawaharada_etal10, walker_etal13, urban_etal14},
deviating from theoretical predictions from hydrodynamical simulations
\citep{tozziandnorman_01,voit_etal03}.  Second, {\em Suzaku} X-ray
measurements of the Perseus cluster revealed an enclosed gas mass
fraction that exceeds the cosmic baryon fraction
\citep{simionescu_etal11}.  Last, X-ray follow-up observations of high
redshift clusters selected by Sunyaev-Zel'dovich (SZ) surveys found
ICM temperatures lower than the theoretical self-similar predictions
in the outer regions of clusters \citep{mcdonald_etal14}.  Several
possible astrophysical phenomena have been proposed to explain the
discrepancy between theoretical predictions and observations,
including gas clumping
\citep{nagaiandlau_11,zhuravleva_etal13,vazza_etal13,roncarelli_etal13}
and non-thermal pressure support \citep[e.g.][]{nelson_etal14b,
  shiandkomatsu14, shi_etal14}.  Another potentially important
astrophysical process that can lead to biased gas measurements in the
outskirts of clusters is the temperature non-equilibrium between
electrons and the heavier ions in the intracluster plasma
\citep{foxandloeb_97,ettoriandfabian_98,wongandsarazin_09,
  ruddandnagai_09}. 

Collisionless shocks convert the bulk kinetic energy of each
  particle species into thermal energy.  Heavier ions possess most of
  the bulk kinetic energy and therefore retain most of the thermal
  energy in the post-shock gas.  The shock heats the electrons to a
  much lower temperature, proportional to the temperature of the
  heavier ions by a factor of the ratio of their masses $m_e/m_i$.
  Electrons are subsequently heated by the ions via Coulomb collisions
  and other interactions.  If Coulomb collisions are the main physical
  mechanism that equilibrates electron and ion temperatures in cluster
  outskirts, equilibration timescales can be very long, leaving
  electrons at colder temperatures than the mean temperature of the
  gas.

Since X-ray and SZ observations are sensitive to the electron
component of the intracluster plasma, non-equilibrium electrons could
introduce a bias in derived physical properties of the ICM in cluster
outskirts.  Derived ICM properties assume that the measured
  electron temperature corresponds to the total thermal energy of the
  plasma.  However, the mean gas temperature, which is an average over
  all particle species, may be higher than the electron temperature.
To understand the origins of the tension between theoretical
predictions and cluster outskirt observations, it is important to
distinguish and quantify the relative importance of non-equilibrium
electrons from other effects.

By analyzing three simulated clusters extracted from cosmological
simulations with the two-temperature effect, \citet{ruddandnagai_09}
showed that the bias in electron temperature is dependent on the mass
and dynamical state of a galaxy cluster.  More massive clusters have
higher temperatures that lead to longer equilibration time scales,
while more disturbed clusters have undergone recent mergers that
generate more non-equilibrium electrons.

In this work, we quantify the temperature bias due to non-equilibrium
electrons using a statistical sample of galaxy clusters from a
high-resolution cosmological simulation of galaxy clusters.  We first
establish and quantify the statistical relationship between the
temperature bias and both the cluster mass and mass accretion rate
(MAR).  By analyzing mock {\em Chandra} observations of simulated 
clusters, we assess the non-equilibrium effects on the ICM temperatures 
measured from observations.  We discuss implications for future X-ray 
and SZ measurements of cluster outskirts.

Our paper is organized as follows. In Section~\ref{sec:methods} we
briefly describe the simulations we used and the mock {\em Chandra}
analysis pipeline.  We present our results in
Section~\ref{sec:results}, and give our summary and discussion in
Section~\ref{sec:summary}.

%%%%%%%%%%%%%%%%%%%%%%%%%%%%%%

%-------------------------------------------------%
\section{Methodology}
\label{sec:methods}
%-------------------------------------------------%

%-------------------------------------------------%
\subsection{Cosmological Simulation with a Two Temperature Model}
%-------------------------------------------------%
We use a mass-limited sample of 65 galaxy clusters from a
high-resolution cosmological simulation {\em Omega500}, the details of
which can be found in \citet{nelson_etal14}.  The simulation box is
$500\,h^{-1}$Mpc on each side, and has a peak spatial resolution of
$3.8\,h^{-1}$kpc. We neglect radiative cooling and star formation, which
should have minimal effects on cluster outskirts.

Using the Adaptive Refinement Tree code \citep{kravtsov_99,
  kravtsov_etal02, rudd_etal08}, we performed the simulation with a
modification to model electrons and heavier ions separately as
described in \citet{ruddandnagai_09}. Electrons and ions are assumed
to be in local thermodynamic equilibrium with separate respective
temperatures $T_e$ and $T_i$, and the relaxation process between the
two components is explicitly calculated.  The electron temperature
time evolution is modeled as,
\begin{eqnarray}\label{eqn:electron_eom}
  \frac{dT_e}{dt}&=&\frac{T_e-T_i}{t_{ei}}-(\gamma-1)T_e\nabla\cdot\mathbf{v},
\end{eqnarray}
where the second term accounts for heating and cooling from adiabatic
accretion, $\gamma = 5/3$ is the adiabatic index, $\mathbf{v}$ is the
gas velocity, and $t_{ei}$ is the equilibration timescale for a fully
ionized medium \citep{spitzer_62} comprised of electrons, protons, and
\ion{He}{2},
\begin{eqnarray}\label{eqn:spitzer_tei}
t_{ei}&=&6.3\times10^8\text{yr}\left(\frac{T_e}{10^7\text{K}}\right)\left(\frac{n_i}{10^{-5}\text{cm}^{-3}}\right)^{-1}\left(\frac{\ln\Lambda}{40}\right)^{-1},
\end{eqnarray}
where $n_i$ is the ion number density, and $\ln\Lambda$ is the Coulomb
logarithm, 
\begin{eqnarray}
  \ln\Lambda&=&37.8+\ln\left(\frac{T_e}{10^7\text{K}}\right)
-\frac{1}{2}\ln\left(\frac{n_i}{10^{-5}\text{cm}^{-3}}\right).
\end{eqnarray}
In this two-temperature model, the hydrodynamic equations are computed
using the total thermal energy of the gas and the electron temperature
is a passively evolving scalar field whose evolution is described by
Equation~\ref{eqn:electron_eom}.  We do not keep track of potential
temperature differences between heavier ions, such as hydrogen and
helium ions, since the equilibration timescales between heavier ions
are very short compared to that of the electrons; i.e., the
equilibration timescale between \ion{He}{2} and protons is a factor of
$\sqrt{m_e/m_{\rm He\,II}}$ of the equilibration timescale between
electrons and protons.

Accretion shocks thermalize the electrons and ions accreting from the
cosmic web. The heavier ions acquire most of the thermal energy, and
heat the electrons through Colulomb collisions in the post-shock
regions. However, in the diffuse cluster outskirts with gas
temperatures around $T>10^7$~K and densities between $10-100$ times
the cosmic mean density, $t_{ei}$ becomes comparable to the Hubble
time, causing the electron temperatures to be lower than the ion
temperatures.

The Spitzer timescale adopted in our simulations is an upper limit of
the true equilibration time, as there are other physical mechanisms
such as plasma instabilities \citep[e.g.,][]{bykov_etal08} that can
couple the temperatures of the different ion species.  However, the
effects of plasma instabilities are expected to be small in the high
Mach number accretion shocks, which are responsible for generating the
non-equilibrium electrons in cluster outskirts.  Plasma instabilities
may provide additional non-adiabatic heating that shortens the
equilibration timescale in gas with lower Mach numbers.  Therefore,
the results in this paper should be taken as a limit on the {\em
  maximal} effects of non-equilibrium electrons.

%%%%%%%%%%%%%%%%%%%%%%%%%%%%%%
\begin{figure*}[htbp]
  \centering
  \mbox{
    \subfigure{\includegraphics[scale=0.5]{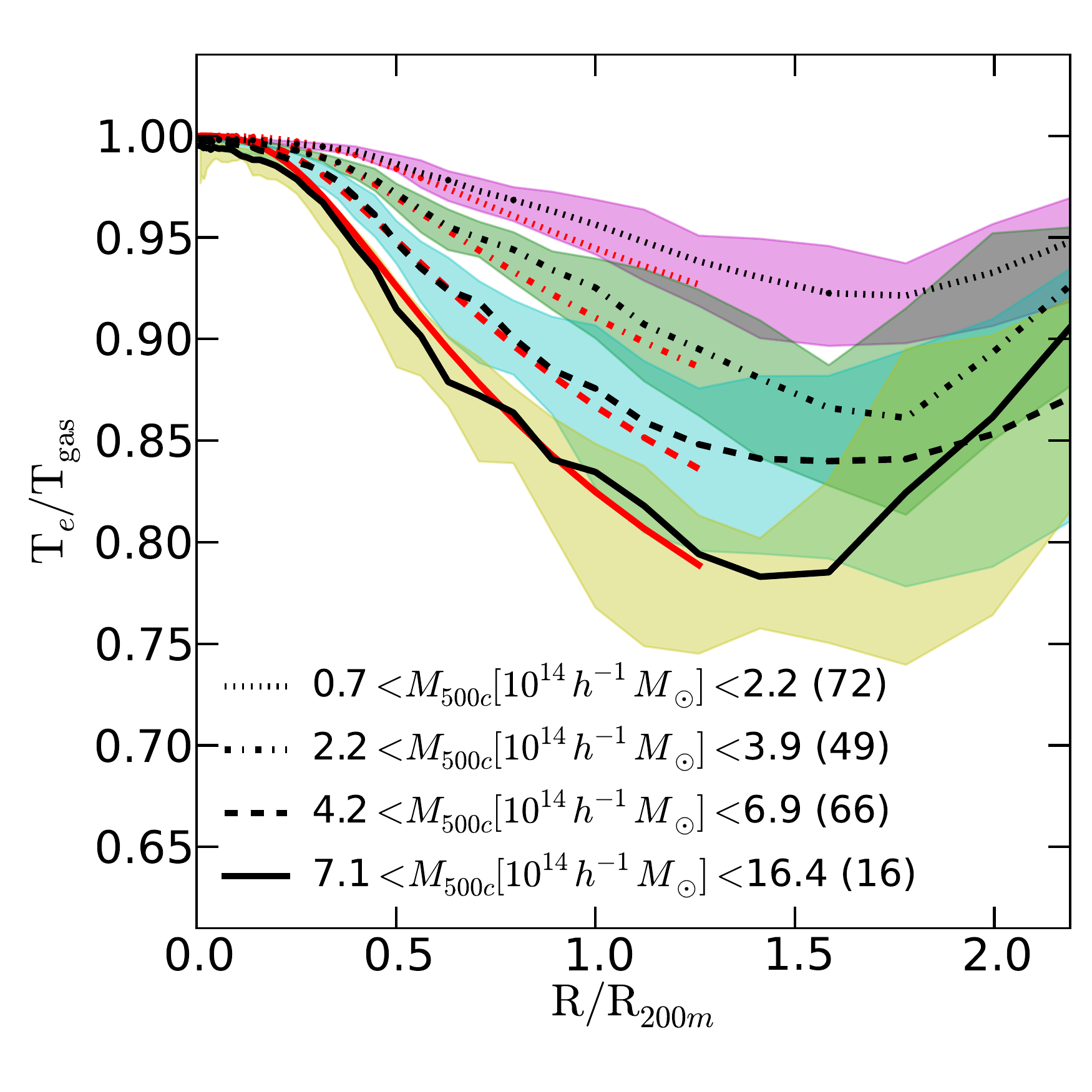}}
    \quad                     
    \subfigure{\includegraphics[scale=0.5]{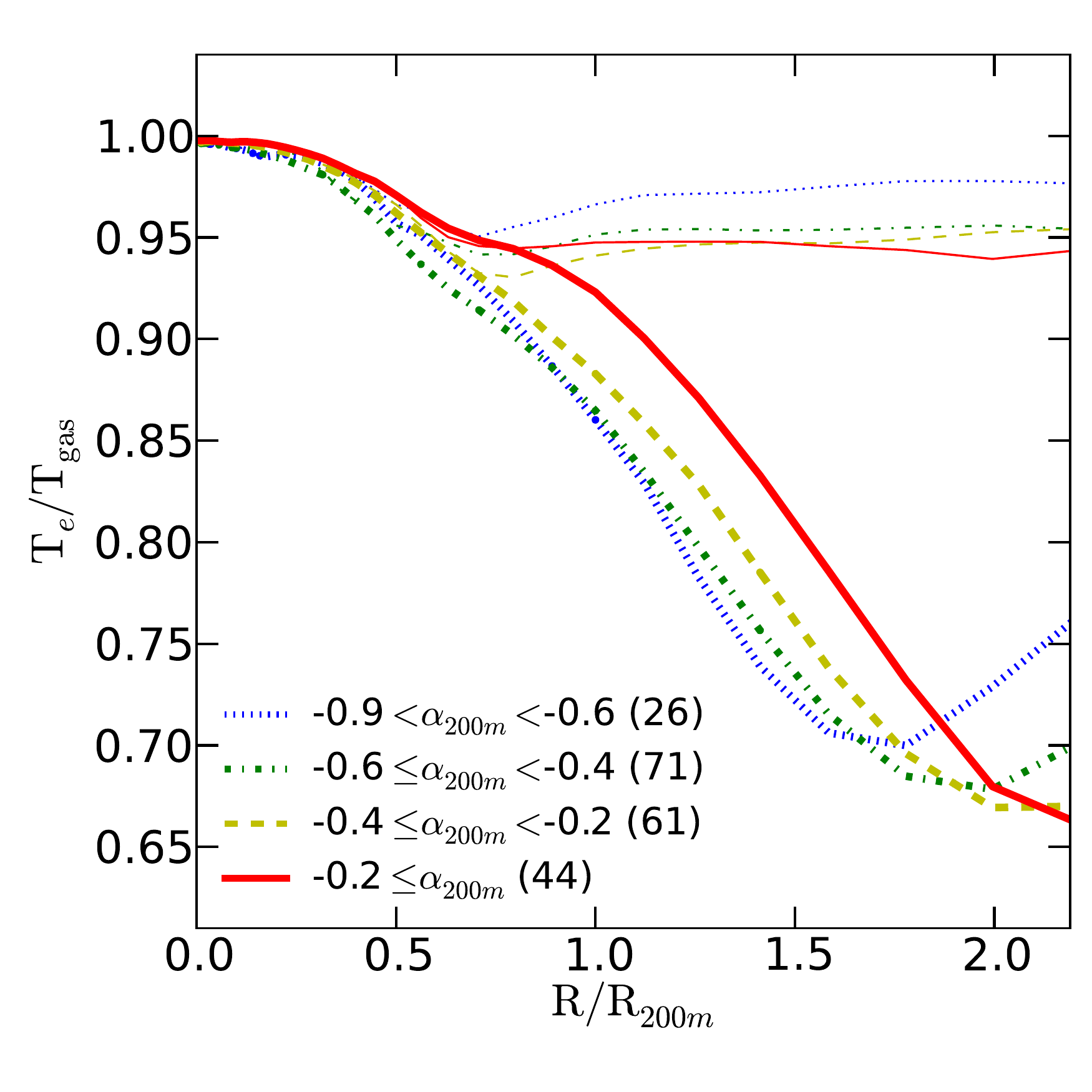}}
  }
  \caption{(Left) Black lines show the average profiles of the
      mass-weighted temperature bias binned by mass.  We have selected
      65 clusters at $z=0$ and their mainline progenitor data at
      $z=0.5$, $1$, and $1.5$ prior to binning the profiles by mass.
    The number of clusters in each mass bin denoted in parentheses in
    the legend label. Red lines show the fitting function
    (Eqn.~\ref{eqn:fitbias}) with input masses that correspond
    to the average mass in each bin. (Right) Thick lines are the
    averaged profiles of the bias in the diffuse ICM, in bins of mass
    accretion rate (MAR) proxy $\alpha_{200m}$; thin lines: the bias
    in the filament component in bins of $\alpha_{200m}$.  Filaments
    exhibit less of a temperature bias due to shorter equilibration
    times and decreased susceptibility to the accretion shock.
    Variations in the amount of recent accretion lead to scatter in
    the self-similar behavior of the temperature bias.  }
  \label{fig:alphamass}
\end{figure*}
%%%%%%%%%%%%%%%%%%%%%%%%%%%%%%

%-------------------------------------------------%
\subsection{Mock {\em Chandra} Analysis}
%-------------------------------------------------%

We create realistic mock X-ray photon maps from X-ray flux maps 
by convolving with {\em Chandra} response files.  We
generate two sets of X-ray flux maps: (1) the first uses a projected
X-ray emissivity that has been generated using the mean gas
temperature in the simulation
$T_{\text{gas}}=(n_iT_i+n_eT_e)/(n_i+n_e)$, where $n_e$ is the
electron density; and (2) the second uses the electron temperature,
$T_e$.

Below we summarize the main elements of the mock {\em Chandra} 
analysis pipeline here. Further details of the pipeline can be found in
\citet{nagai_etal07b} and \citet{avestruz_etal14}. 

The X-ray emissivity for a given $k$-th hydrodynamical cell with volume
$\Delta V_k$ in the simulation is given by,
\begin{eqnarray}
j_{E,k}= n_{e,k}n_{i,k}\Lambda_{E}(T_k,Z_k,z)\Delta V_k,
\end{eqnarray}
where $n_e$ and $n_i$ are the respective number densities of electrons
and ions, $T_k$ is either the mean gas temperature ($T_{\text{gas}}$)
or the electron temperature of that element ($T_e$), $Z_k=0.3Z_\odot$
is the assumed metallicity, and $z$ is the redshift.  We assume a
constant metallicity that is representative of a plausible average
metallicity for a galaxy cluster, since our simulation does not follow
cooling and star formation. We compute the X-ray emissivity,
$\Lambda_E(T_k, Z_k, z)$, using the MEKAL plasma code
\citep{mewe_etal85,kaastra_etal93,liedahl_etal95}.  We multiply the
plasma spectrum by the Galactic absorption corresponding to a
hydrogen column density of $N_{\mathrm{H}}=2\times 10^{20}$~cm$^{-2}$.

We convolve the emission spectrum with the response of the {\em
  Chandra} ACIS-I CCDs and draw photons from each position and spectral
channel according to Poisson distribution.  The resulting photon
maps have an exposure time of $2.4$~Msec, similar to the deep {\em
  Chandra} observations of Abell 133 (Vikhlinin~et~al., in prep.), 
and comparable in photon counts
to stacked galaxy cluster analyses.  From images generated in the
$0.7-2$~keV band, we identify and mask out clumps using the wavelet
decomposition algorithm described in \citet{vikhlinin_etal98}.  We
also exclude regions that correspond to overdense filamentary
structures, the details of which are described in \citet{avestruz_etal14}.

%-------------------------------------------------%
\section{Results}
\label{sec:results}
%-------------------------------------------------%

%-------------------------------------------------%
\subsection{Dependence on mass and mass accretion rate}
\label{sec:massdep}
%-------------------------------------------------%

The temperature bias from non-equilibrium electrons depends on
  the location of the accretion shock that generates the initial
  temperature difference between electrons and heavier ions and the 
  relative age of the gas ($\sim M/\dot{M}$) compared with the equilibration 
  timescale in Equation~\ref{eqn:spitzer_tei}. The
former depends on the mass accretion rate (MAR) of the cluster, and
the latter depends on the the temperature and density (hence the mass)
of the cluster.

The left panel of Figure~\ref{fig:alphamass} shows the profiles of the
electron temperature bias $T_e/T_{\rm gas}$ of the simulated clusters
at $z=0,0.5,1,1.5$ averaged in four mass bins, with masses
  defined within ${R}_{200m}$ (black lines). Note that we have
  taken the profiles of the 65 clusters at $z=0$, $0.5$, $1$, and
  $1.5$ prior to binning the profiles by mass.

In a companion paper \citep{lau_etal14}, we show that the average
location of the accretion shock occurs at a fixed fraction of
$R_{200m},$\footnote{$R_{200m}$ is the cluster radius enclosing an
  average density $200$ times the mean matter density of the
  universe.} independent of redshift.  The location of the maximum
temperature bias is mainly driven by the accretion shock, and
therefore also exhibits the same redshift independence when scaled to
${R}_{200m}$.  The $T_e/T_{\rm gas}$ profile for each mass bin shows a
consistent location for the maximum temperature bias, which is located
at the accretion shock radius $R_{\rm sh} \approx 1.6\times R_{200m}$.
At this radius, the shocked electrons are maximally out of equilibrium
with the heavier ions.  The extent of the bias is larger for more
massive clusters with hotter gas and longer equilibration times (see
Equation~\ref{eqn:spitzer_tei}).

The electron temperature bias also depends on the MAR of the cluster.
The MAR shifts the location of the accretion shock that generates the
non-equilibrium electrons. We use a proxy of the MAR of the cluster
\begin{equation}\label{eqn:alpha}
\alpha_{200m} \equiv V_r^{DM} (r=R_{\alpha})/V_{{\rm circ},200m},
\end{equation}
defined as the average mass-weighted dark matter radial
velocity $V_r^{DM}$ measured at radius $R_{\alpha} = 1.25 R_{200m}$
and normalized by the circular velocity $V_{{\rm circ},200m} \equiv
\sqrt{GM_{200m}/R_{200m}}$ of the cluster \citep{lau_etal14}.  A more
negative $\alpha_{200m}$ indicates higher mass accretion rate.

The value of $R_{\alpha}$ was chosen to minimize scatter relative to an
  alternative mass accretion rate proxy \citep{diemerandkravtsov_14}, 
 \begin{equation}
  \Gamma_{200m}=\frac{\log_{10}(M_{200m}(z=0)/M_{200m}(z=0.5))}{\log_{10}(a(z=0)/a(z=0.5))}
 \end{equation}
  which only describes the mass accretion rate of a cluster at $z=0$
  \citep[see Figure~5 in][]{lau_etal14}. The choice of a fixed fraction
  of ${R}_{200m}$ to measure an accretion rate provides a consistent
  way to compare accretion rates of clusters at different epochs.
 
The right-hand panel of Figure~\ref{fig:alphamass} shows the profile
of the temperature bias in bins of $\alpha_{200m}$.  Following
\citet{lau_etal14}, we decompose gas in a given radial bin into
``filament'' and ``diffuse'' components according to the density and
radial velocity of the gas defined with respect to the cluster center.  At
large cluster-centric radii, a large fraction of the gas mass belongs to
dense filaments and clumps, with net radial velocities that point
towards the cluster center.  The thick lines in the plot correspond to
the bias in the diffuse ICM of the clusters, and the thin lines
correspond to the bias in the filaments. 

In the filament component of the gas (thin lines), the electron
  temperature is very close to the mean gas temperature.  Gas in
  filaments also has higher densities and lower temperatures than gas
  in the diffuse component, leading to shorter equilibration times.
  In the diffuse component of the gas (thick lines), the location of
  maximum bias occurs closer to the cluster center for more rapidly
  accreting clusters.  Clusters with higher accretion rates have more
  negative values of $\alpha_{200m}$.  Accreted material in these
  clusters have higher momentum fluxes, causing the location of
  accretion shock to move inward.

\begin{figure*}[htbp]
  \vspace{15pt}
  \centering
  \mbox{
    \subfigure{\includegraphics[scale=0.85, trim=0 0 1.5cm 0, clip=True]{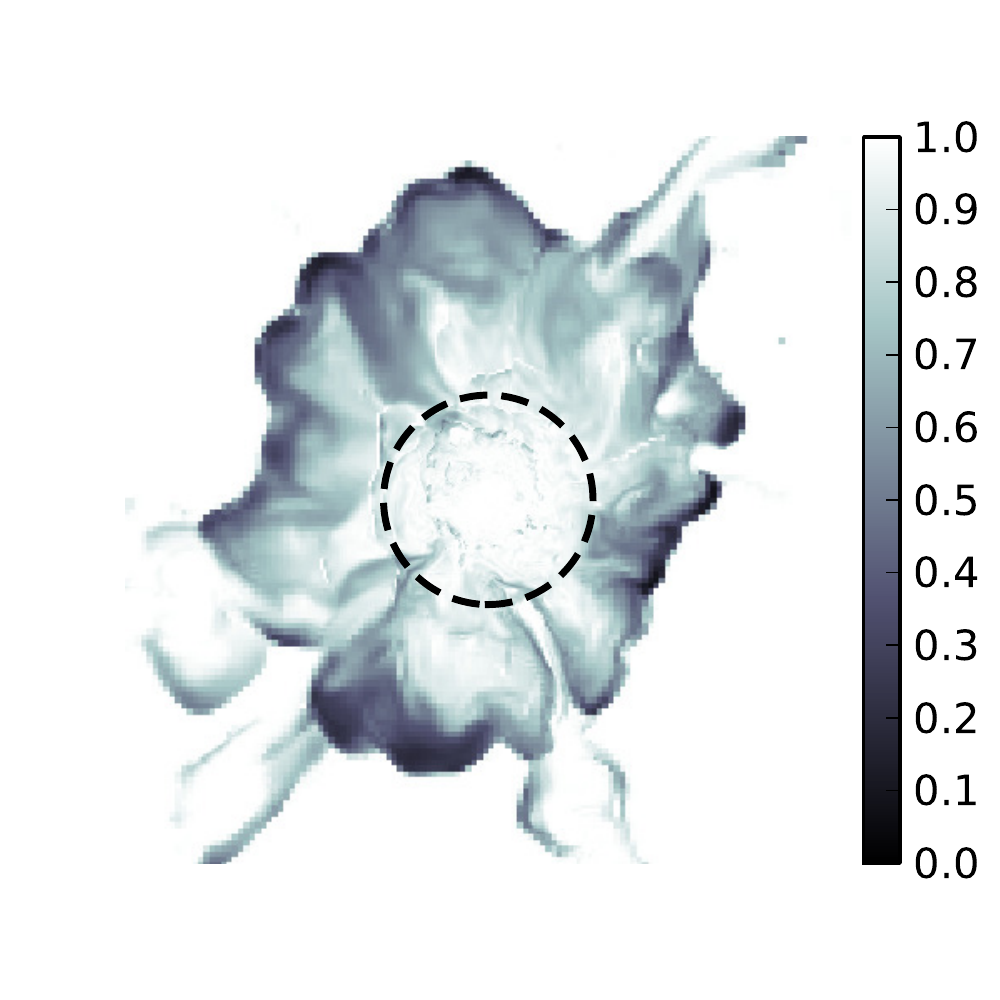}}\quad                     
    \subfigure{\includegraphics[scale=0.85, trim=0 0 0 0, clip=True]{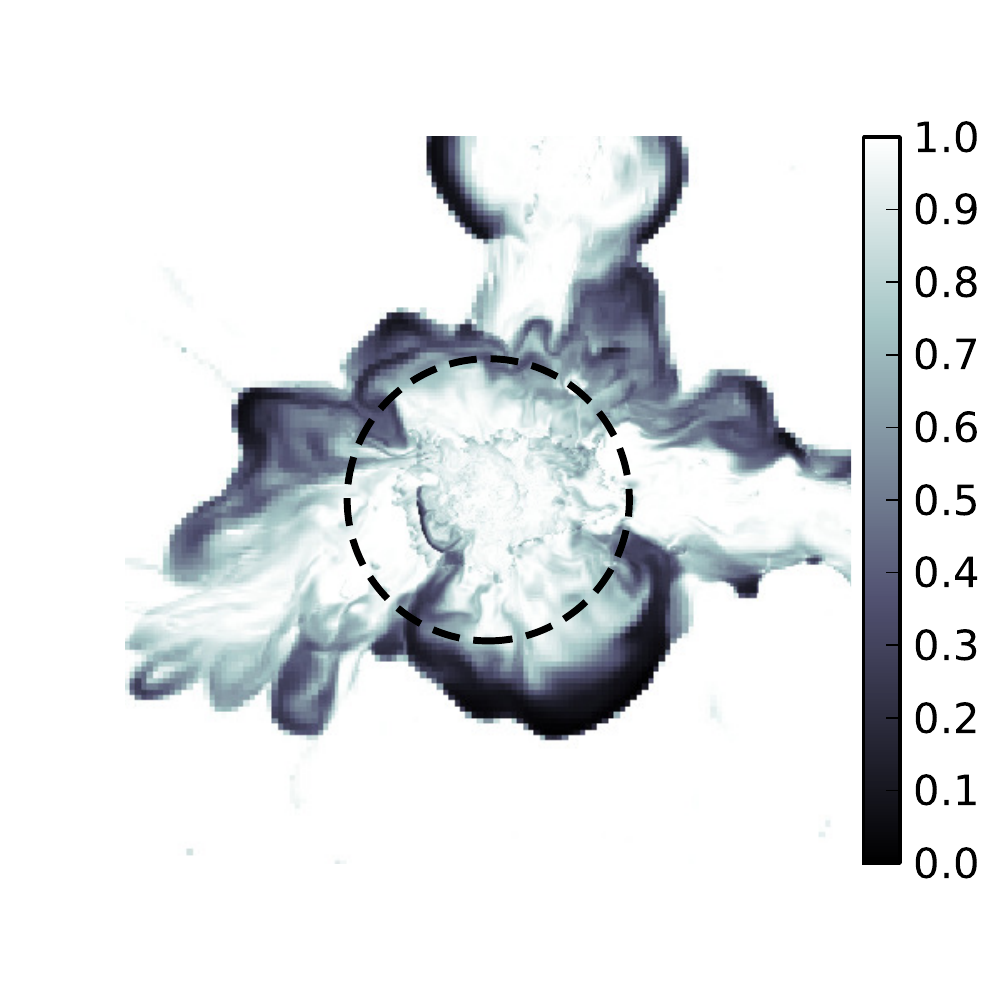}}
  }
  \caption{Sliced maps of $T_e/T_{\rm gas}$ for the most relaxed and
    least massive cluster (left panel), and the most massive and least
    relaxed cluster (right panel) from our simulation.  The colorbar
    indicates $T_e/T_{\rm gas}$.  The dimension for each panel is
    $15.6\,h^{-1}\Mpc\times 15.6\,h^{-1}\Mpc$, with depth of
    $7.6\,h^{-1}\kpc$.  The circle in dashed line shows $R_{200m}$ of
    the cluster.  Between $1.0<R/R_{200m}<1.5$, $T_e/T_{\rm gas}$ is
    close to unity in the filamentary gas with high momentum flux
    entering the cluster well, but is significantly less than unity in
    the diffuse gas at the same cluster-centric radii.  The right
    panel shows gas from the bottom-left filament experiencing an
    accretion shock at radii smaller than $R_{200m}$.}
  \label{fig:tetgasmap}
\end{figure*}

The distinction between the diffuse ICM and the filaments is visible
in Figure~\ref{fig:tetgasmap} where we show slices of 
$T_e/T_{\rm gas}$ for a relaxed low mass cluster (left panel) and a
merging massive cluster (right panel) from our simulation sample.  The
dashed circles indicate ${R}_{200m}$ for each cluster.  Darker shades
in the maps indicate regions where more electrons are out of
equilibrium with the heavier ions, corresponding to a larger
temperature bias.  The regions of larger bias mostly correspond to gas
in the diffuse ICM.  The lighter shades indicate regions where the
electron temperature is very close to the mean gas temperature.  Near
$R_{200m}$, the lighter shades correspond to gas in filaments, which
experience shock heating at smaller radii than gas in the diffuse ICM.

In the more massive and less relaxed cluster shown in the right panel,
mergers drive more small scale shocks that lead to more
non-equilibrium electrons.  Additionally, the higher ICM temperature
in the outskirts of the more massive cluster leads to longer
equilibration times and therefore larger temperature biases.

%%%%%%%%%%%%%%%%%%%%%%%%%%%%%%
%%%%%%%%%%%%%%%%%%%%%%%%%%%%%%

The temperature bias in the left panel of Figure~\ref{fig:alphamass}
uses the mass-weighted temperatures from {\it all} of the cluster gas.
Since filaments are denser in cluster outskirts, the temperature bias
is more heavily weighted towards the filament contribution, where the
non-equilibrium effect is small.

%-------------------------------------------------%
\subsection{Redshift dependence}\label{sec:zdep}
%-------------------------------------------------%

%%%%%%%%%%%%%%%%%%%%%%%%%%%%%%
\begin{figure}[t]
  \centering
  \includegraphics[scale=0.45]{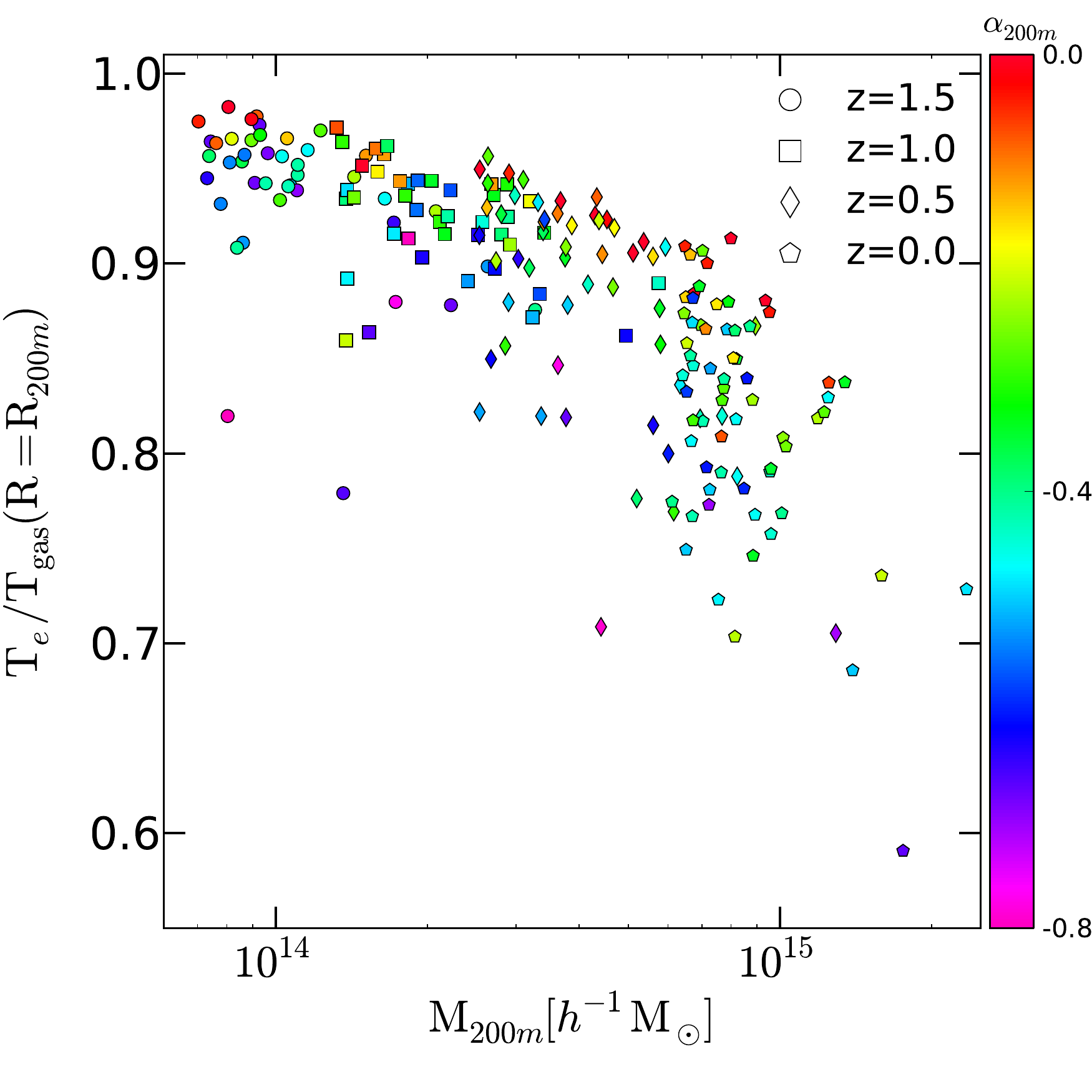}
  \caption{The temperature bias of non-equilibrium electrons as a
    function of cluster mass, both measured at $r={R}_{200m}$.  We show
    the bias for all 65 clusters at $z=0$, $0.5$, $1$, and $1.5$, with
    respective marker styles of circles, squares, diamonds, and
    triangles.  For each data point, we color code the value of the
    mass accretion rate parameter, $\alpha_{200m}$. 
    }
  \label{fig:tetgasr200m}
\end{figure}
%%%%%%%%%%%%%%%%%%%%%%%%%%%%%%

Figure~\ref{fig:tetgasr200m} shows $T_e/T_{\text{gas}}$ as a function
of cluster mass, both measured at $r={R}_{200m}$.  Different redshifts
($z=0,0.5,1.0,1.5$) are indicated by marker styles (respectively
circles, squares, diamonds, and triangles).  There is an apparent
redshift dependence of $T_e/T_{\text{gas}}$, with high-$z$ clusters
exhibiting less of a bias than low-$z$ clusters.  However, the
redshift dependence is primarily due to cluster mass growth.  At low
redshift, there are more high mass clusters with a more pronounced
bias.  For a given cluster mass, there is no significant trend of
$T_e/T_{\text{gas}}$ with redshift.  There is, however, a systematic
trend in $T_e/T_{\text{gas}}$ with MAR.  Clusters with higher MAR
(more negative values of $\alpha_{200m}$, as indicated by the color of
the points in Figure~\ref{fig:tetgasr200m}) tend to a larger bias,
with lower $T_e/T_{\text{gas}}$ values at ${R}_{200m}$.  As discussed
in Section~\ref{sec:massdep}, clusters with higher MAR have accretion
shocks closer to the cluster center, leading to an inward shift
  of the location of maximum bias.

We note that there can be additional redshift dependence in
$T_e/T_{\text{gas}}$ due to any evolution in the distribution of
  cluster assembly history.  At a given redshift, more recently
  assembled clusters have more recently accreted electrons with less
time to equilibrate; we expect the fraction of recently assembled
  clusters to vary with redshift.  The mass accretion timescale
$t_{\rm acc}\equiv{M}/\dot{M}$ for a given cluster at a given redshift
is an indicator of cluster age.  Our MAR proxy, $\alpha_{200m}$
(defined in Equation~\ref{eqn:alpha}), is inversely proportional to
$t_{\rm acc}$, as $V_{\rm circ,200m}$ is a proxy for cluster mass and
$V_r^{DM}(r=R_{\alpha})$ describes the instantaneous MAR.  However, in
our sample, there does not appear to be any strong correlation between
$z$ and $\alpha_{200m}$ for a given mass.  To make a more stringent
statement on any additional redshift dependence, we would need a
larger box that samples a wider mass range in a given redshift bin.

%-------------------------------------------------%
\subsection{Fitting function of the temperature bias}
\label{sec:fit}
%-------------------------------------------------%

We provide a simple fitting function for the profile
of the temperature bias:
\begin{eqnarray}\label{eqn:fitbias}
b_e(R)&\equiv& \frac{{T_e}(R)}{{T_{\text{gas}}(R)}} = \frac{(x/x_t)^{-a}}{\left(1+(x/x_t)^{-b}\right)^{a/b}},
\end{eqnarray}
where $x=R/R_{200m}$ and the three best-fit parameters that account for the dependence on cluster mass are given by,
\begin{eqnarray}
  x_t&=&0.629\times\left({M}_{200m}/10^{14}h^{-1}{M}_{\odot}\right)^{-0.1798} \nonumber \\
  a&=&0.086\times\left({M}_{200m}/10^{14}h^{-1}{M}_{\odot}\right)^{0.3448}\\
  b&=&2.851\times\left({M}_{200m}/10^{14}h^{-1}{M}_{\odot}\right)^{-0.006}. \nonumber
\end{eqnarray}
In the left panel of Figure~\ref{fig:alphamass}, we overplot profiles
of the fitting function from Equation~\ref{eqn:fitbias} in red. Each
red line corresponds to the fitting function with an input mass
  that is the average cluster mass in each cluster mass bin.  The
fitting function adequately describe the averaged profiles of the bias,
shown as black lines.

Here we do not decompose the ICM into diffuse and filament components,
nor do we separately model dependence on $\alpha_{200m}$, since
applications of the fitting function likely treat spherical averages
of the halo profile and do not typically have the information
necessary to calculate $\alpha_{200m}$.  The dependence on
$\alpha_{200m}$ comes from the fact that accretion rate shifts the
location of the shock radius.  We have characterized the relationship
between the shock radius and accretion rate in Figure~10 of the
companion paper by \citet{lau_etal14}, which shows that the scatter in
the shock radius at a given is $\alpha_{200m}$ nearly a factor of
two. This large scatter in the shock radius stems from highly
aspherical and complex geometry at large radii.  As a result, the
location of the shock radius exhibits a relatively weak dependence
(although systematic) on $\alpha_{200m}$, whereas the mass dependence
of the temperature bias exhibits a stronger behavior with better
predictive power. We therefore use the fitting function to describe
the bias in a statistical manner, parameterizing by mass alone.

%-------------------------------------------------%
\subsection{ICM profiles from mock X-ray maps}
\label{sec:icmprofiles}
%-------------------------------------------------%

X-ray measurements of the ICM are only sensitive to the electron component
of the plasma.  A temperature bias due to non-equilibrium electrons
propagates to ICM quantities inferred from the X-ray temperature.

We measure the X-ray temperature profiles from two sets of mock {\em Chandra} 
maps: one generated with the mean gas temperature, and the other with the electron
temperature.  
This allows us to compare the temperature difference 
accounting for instrumental response, projection effects, 
and the spectroscopic weighting of the X-ray gas.

\begin{figure}[t]
  \centering
  \includegraphics[scale=0.45]{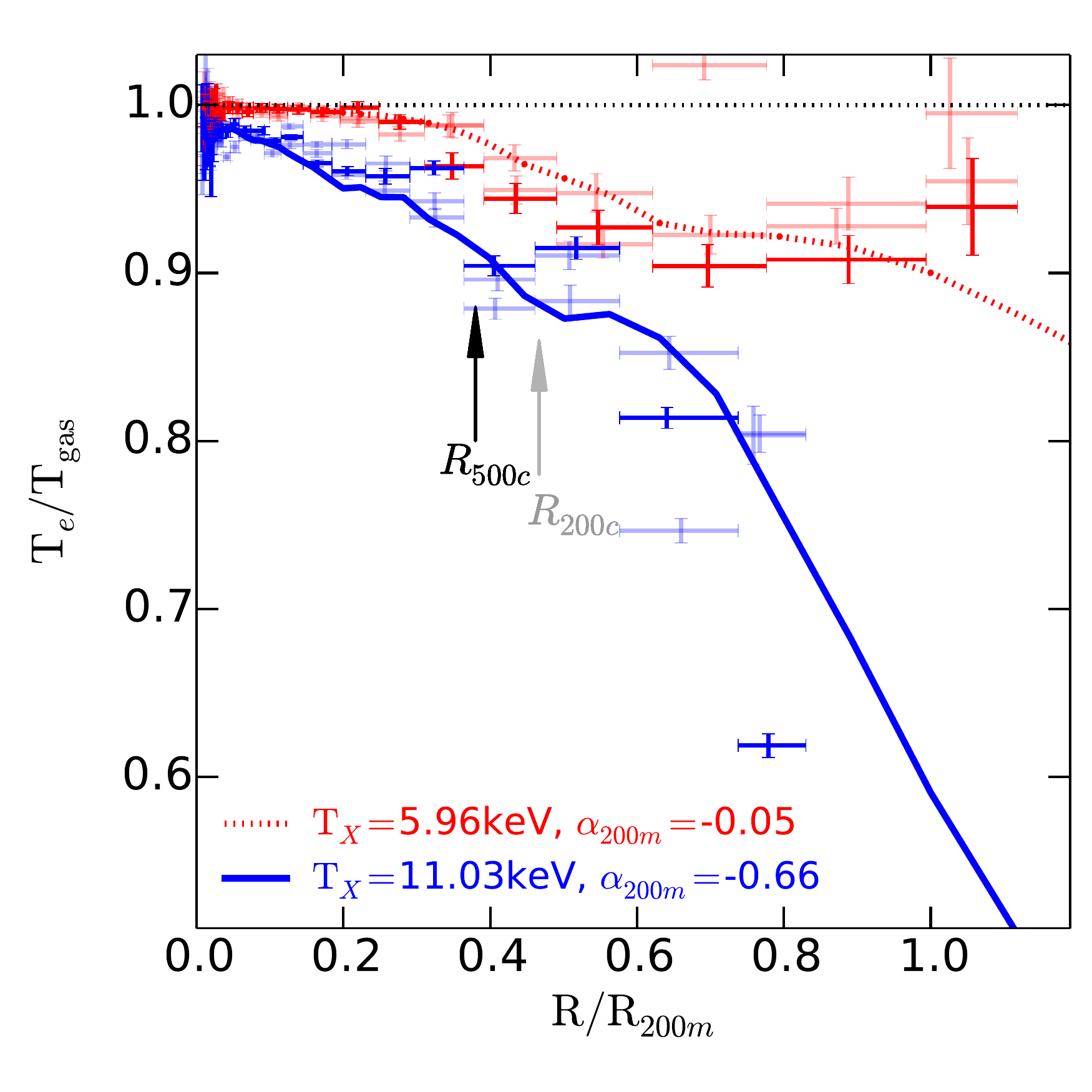}
  \caption{Lines show the temperature bias profiles of the diffuse ICM
    in one of our most massive and least relaxed clusters (blue
    solid), and one of our least massive and most relaxed clusters
    (red dotted).  Data points show the corresponding bias of the
    diffuse ICM measured from mock {\em Chandra} X-ray maps.  The
      light red and blue data points correspond to the respective
      temperature biases of the least massive and most relaxed cluster
      along two other orthogonal projections to illustrate line-of-sight
      variations.  The black and grey arrows denote ${R}_{500c}$ and
    ${R}_{200c}$ respectively. }
  \label{fig:txetxgas}
\end{figure}

Figure~\ref{fig:txetxgas} shows the bias in the X-ray projected
temperature profile, measured from spectral fitting of photons from
each of the two mock maps.  The bias is then calculated as a ratio of
measured X-ray electron temperature to the X-ray temperature that
would be measured if electrons were in equilibrium with heavier ions.
Red data points show the projected X-ray temperature bias for one of
the less massive relaxed clusters with $T_X=5.96$ keV
($M_{200m}=7.2\times10^{14}h^{-1}M_{\odot}$) and
$\alpha_{200m}=-0.05$.  Blue data points show the same bias in one of
our more massive unrelaxed clusters with $T_X=11.03$ keV
($M_{200m}=1.75\times10^{15}h^{-1}M_{\odot}$) and
$\alpha_{200m}=-0.66$.  These two representative clusters bracket the
effects of non-equilibrium electrons on X-ray measurements in this
mass range, but represent upper limits of the temperature bias in both
cases.  For the relaxed less massive cluster, the bias within
$R<{R}_{500c}$ is less than $5\%$.  In the unrelaxed, more massive
cluster that has experienced rapid recent accretion, the bias is
$\sim10\%$ at ${R}_{500c}$, and increases precipitously at larger
radii.

We have also overplotted the ratio of the mass-weighted electron
temperature profile to the mass-weighted mean temperature profile of
the diffuse ICM, computed directly from the simulation. The agreement
between the bias in mock X-ray measurements and the bias in the
mass-weighted temperature shows that the bias is comparable between
the two methods of measuring the ICM temperature.

Note that the upturn in the bias at ${R}\approx0.8{R}_{200m}$
for the red mock X-ray data in the less massive cluster is due to
line of sight effects; the bias is not perfectly spherical, even in
the bulk component.  Some projections will have a bias in the
projected X-ray temperature that is either below or above the bias
calculated from the spherically averaged mass-weighted temperatures
shown in the red dotted line.  In this projection, there is slightly
less shock heated material at this radius compared to other lines of
sight.

Likewise, the drop in the X-ray temperature bias for the more massive
less relaxed cluster at ${R}\approx0.8{R}_{200m}$ is also due to the
line of sight; the line of sight for the X-ray data has more shock
heated material at that radius compared to other lines of sight, all
of which have a mass-weighted average shown in the solid blue line.

To illustrate the scatter in projected temperature biases, we
  have also plotted, in lighter colors, the measured X-ray temperature bias 
  along two other orthogonal projections for each cluster.  
  In both clusters, the temperature biases are generally present in all three 
  projections. However, the temperature measurements from X-ray mocks 
  occasionally include photons from clumps and filaments that were not completely 
  masked out; these photons can contribute to the projected annular average by 
  driving the temperature ratio closer to unity, since the high-density gas in 
  clumps and filaments has less of a temperature bias.

%%%%%%%%%%%%%%%%%%%%%%%%%%%%%%
%%%%%%%%%%%%%%%%%%%%%%%%%%%%%%

%-------------------------------------------------%
\subsection{The effects on hydrostatic mass estimates and scaling relations}
\label{sec:hsescale}
%-------------------------------------------------%

%--------------------------------------------

\begin{table}
\caption{Biases in hydrostatic mass and gas temperature of galaxy clusters at $z=0$}
\begin{center}
\begin{tabular}{cccc}
\hline \\ [-0.2cm]
$T_X$ [keV] & $R=R_{500c}$ & $R=R_{200c}$ & $R=R_{200m}$ \\ [0.2cm]
\hline \\ [-0.2cm]
\multicolumn{4}{c}{(a) $M_{\rm HSE}(R;T_e)/M_{\rm HSE}(R;T_{\rm gas})$} \\ [0.2cm]
\hline \\ [-0.2cm]
4.2--6.1 (38) & $0.993 \pm 0.002$ & $0.959 \pm 0.007$ & $0.889 \pm 0.009$ \\ 
6.1--8.7 (24) & $0.989 \pm 0.005$ & $0.926 \pm 0.010$ & $0.868 \pm 0.012$ \\ 
8.7--12.5 (3) & $0.961 \pm 0.009$ & $0.893 \pm 0.016$ & $0.803 \pm 0.041$  \\ [0.2cm]
\hline \\ [-0.2cm]
\multicolumn{4}{c}{(b) core excised $b_e(<R) \equiv T_e(<R)/T_{\rm gas}(<R)$} \\ [0.2cm]
\hline \\ [-0.2cm]
4.2--6.1 (38) & $0.991 \pm 0.001$ & $0.985 \pm 0.002$ & $0.975 \pm 0.001$ \\ 
6.1--8.7 (24) & $0.985 \pm 0.002$ & $0.976 \pm 0.003$ & $0.967 \pm 0.003$ \\ 
8.7--12.5 (3) & $0.972 \pm 0.003$ & $0.962 \pm 0.003$ & $0.952 \pm 0.004$  \\ [0.2cm]
\hline
\end{tabular}
\end{center}
\label{tab:bias}
\tablecomments{Panel (a) shows the mean biases in hydrostatic mass and their 
$1\sigma$ errors due to non-equilibrium electrons at the given radius $R$ . 
Panel (b) shows the mean bias the mass-weighted temperature integrated 
between $0.15 R_{500c}$ and $R$ specified in the first row. 
The number in the parentheses indicate the number of clusters in each bin of $T_X$. }
\end{table}

We calculate the corresponding effect of non-equilibrium electrons on
hydrostatic mass estimates for our cluster sample in four temperature
bins.  In Table~\ref{tab:bias}, we list the ratios between the
hydrostatic mass estimates calculated with electron temperature to
that with the mean gas temperature.  The hydrostatic mass estimate
from the electron temperature corresponds to the observationally
inferred mass in the presence of non-equilibrium electrons, while the
hydrostatic mass estimate from the mean gas temperature corresponds to
the inferred mass with electrons fully in equilibrium with the heavier
ions.

In the most massive clusters ($T_X\gtrsim8$~keV), the core-excised mass-weighted 
electron temperature is biased low by $3\%$ at $R_{500c}$, and as much as $5\%$ at
${R}_{200m}$.  The corresponding bias in the observed hydrostatic mass
for these clusters is $4\%$ at ${R}_{500c}$, and $20\%$ at
${R}_{200m}$.  The bias in observed hydrostatic mass is larger than
the bias in the temperature, $b_e(R)$, at each of these radii, as the
hydrostatic mass within a radius is proportional to both the
temperature and the temperature gradient at that radius.

Table~\ref{tab:bias} also lists the resulting biases in mass-weighted
cluster temperature, $b_e(<R)\equiv T_e(<R)/T_{\rm gas}(<R)$, measured
out to three different values of $R$: $R_{500c}$, $R_{200c}$, and
${R}_{200m}$.  $b_e(<R)$ is an core-excised integrated quantity
between the $0.15R_{500c}$ and the maximum given radius, $R$.
$b_e(<R)$ have a smaller bias than the temperature bias $b_e(R)$
measured at $R$ since the temperature bias monotonically increases
until the shock radius.  The bias in the core-excised temperature is
$<3\%$ at ${R}_{500c}$ and $<5\%$ at ${R}_{200m}$, smaller than the
biases in both the electron temperature at those radii and the
observed hydrostatic mass.

Since the temperature bias due to non-equilibrium electrons has a
larger effect on the hydrostatic mass estimate than on the integrated
temperature, scaling relations such as the $M-T_X$ and $M-Y_X$ will
have a shallower slope with increasing cluster mass, leading to
deviations from self-similarity.

%-------------------------------------------------%

\section{Summary and Discussion}
\label{sec:summary}

%-------------------------------------------------%

We have used a mass-limited sample of 65 galaxy clusters simulated in
a cosmological volume to characterize the effects of non-equilibrium
electrons on gas temperature in the outskirts of galaxy clusters.  In
this work, we quantified the dependence of the temperature bias 
$T_e/T_{\text{gas}}$ on cluster mass and dynamical state in both 
the diffuse and filamentary components of the ICM in cluster outskirts. 
We provided a mass-dependent fitting function for $T_e/T_{\text{gas}}$, 
which will be useful in quantifying the astrophysical uncertainties due 
to non-equilibrium electrons, and in placing constraints on physical processes in
accretion shocks.  Additionally, we used mock {\em Chandra} X-ray maps
with and without non-equilibrium electrons
to determine how the temperature bias manifests in ICM profiles derived from X-ray
observations.

To summarize our findings: 

%%%%%%%%%%%%%%%%%%%%%%%%%%%%%%%%%%%%%%
\begin{itemize}

\item The temperature bias from non-equilibrium electrons in the diffuse
  component of the ICM can reach $40\%$ within $R\leq R_{200m}$, and is
  dependent on the mass and mass accretion rate of the cluster.  More massive
  clusters have longer equilibration times due to their higher gas
  temperatures, leading to a more strongly biased electron temperature
  at all radii. Clusters with higher mass accretion rates have smaller
  accretion shock radius, leading to stronger bias at any given radius
  inside the shock.
     
\item The bias is azimuthally asymmetric.  Due to lower temperatures,
  higher densities, and a net momentum flux towards the cluster
  center, the effects of non-equilibrium electrons in gas filaments is
  much smaller than in the diffuse component of the ICM.

\item Non-equilibrium electrons can affect X-ray measurements of
  galaxy cluster outskirts by biasing the projected X-ray temperatures low.
  The magnitude of the temperature bias accounting for
  instrumental response, projection effects, and spectroscopic
  weighting is similar to the temperature bias measured directly from
  simulations.

\item Non-equilibrium electrons introduce biases in the hydrostatic mass estimates 
as well as the average cluster temperature of massive clusters. In the
  hottest ($T_X\gtrsim8$~kev) clusters, the hydrostatic mass is biased
  low by $4\%$ at ${R}_{500c}$ and $20\%$ at $R_{200m}$, while the
  core-exicsed mass-weighted temperature is biased low by $3\%$ and
  $5\%$, respectively.  Since the bias in the hydrostatic mass estimates
  is larger than the bias in the average cluster temperatures and the biases 
  are larger for hotter clusters, the presence of non-equilibrium electrons leads 
  to shallower slopes in the high mass end of the $M-T_X$ and $M-Y_X$ scaling 
  relations than self-similar predictions.

\item We provide a mass dependent fitting function for the
  profile of the temperature bias (Equation~\ref{eqn:fitbias}). The
  fitting function can be useful in the theoretical modeling of electron-ion
  equilibration in galaxy clusters.
   
\end{itemize}
%%%%%%%%%%%%%%%%%%%%%%%%%%%%%%%%%%%%%%

Our results have implications for X-ray and SZ observations of cluster
outskirts.  The temperature bias due to non-equilibrium electrons
leads to underestimates in  entropy, pressure, and hydrostatic mass
that are of the same size as the temperature bias.  Since gas entropy 
is directly proportional to gas temperature, the temperature bias of order
$\lesssim 15\%$ at $r \gtrsim R_{200c}$ can partially explain the flattening 
of entropy observed by {\em Suzaku}, in addition to gas clumping that biases 
the density high and entropy low. 
The effects of non-equilibrium electrons must also be considered 
when interpreting the recent {\em Planck} stacked
SZ measurements of electron pressure profiles, which measured the
pressure profiles out to $3 \times R_{500c} \approx 1.2 R_{200m}$
\citep{planck_interV2013}.  The effect of non-equilibrium electrons is
mass dependent; the corresponding temperature bias could lead to
deviations from self-similarity in mass-observable scaling relations.

The fitting function provided in Equation~\ref{eqn:fitbias} can also be
useful in bracketing the effects of non-equilibrium electrons on
measurements of the SZ power spectrum \citep[][]{hillandpajer_13} and
SZ bispectrum \citep{hillandsherwin2013,crawford_etal2014}. The
temperature bias will be especially significant in the SZ bispectrum,
since the bispectrum is most sensitive to massive clusters at low
redshift \citep{bhattacharya_etal2012}, whose outskirts would have the
largest temperature bias due to non-equilibrium electrons.

Our quantitative study of non-equilibrium electrons in galaxy cluster
outskirts assumes a maximal equilibration Spitzer timescale. Our model
assumes negligible electron heating within shocks, but there can be
non-adiabatic heating due to plasma instabilities
\citep[e.g.,][]{bykov_etal08}. However, these are expected to be small
in the accretion shocks whose Mach numbers are high.  The results 
presented in this work thus provide an {\it upper} limit of the temperature 
differences between ions and electrons.  

While the temperature bias in high Mach number accretion shocks
  is not directly measurable from current observational data, there
  has been some observational evidence indicating a rapid
  equilibration timescale in low Mach number merger shocks by
  measuring the width of the merger shock in colliding galaxy clusters
  \citep[e.g.,][]{markevitch06}.  The equilibration timescale can be
  constrained by measuring the temperature jump across the shock.  An
  equilibration timescale that is significantly shorter than the
  Coulomb timescale will lead to a sharper electron temperature jump,
  more similar to the temperature jump of the heavier ions.  A similar
  observational test is possible for accretion shocks in cluster
  outskirts.  However, accretion shocks are too faint to detect with
  the current generation of X-ray telescopes, and such measurements
  will not be possible until the next generation of X-ray missions.

Since non-equilibrium electrons have a progressively larger effect on
ICM temperature measurements at increasing cluster-centric radii,
non-equilibrium physics can have an impact on the electron pressure
profiles in clusters outskirts that have recently become accessible to
SZ observations, e.g, with {\em Planck}.  Detailed comparisons of
observed and simulated pressure profiles can therefore provide further
observational constraints on the equilibration efficiency between
electrons and protons.

Finally, one can also assess the non-equilibrium state of other ions
by measuring line intensity ratios of different metals with
high-resolution X-ray spectroscopy onboard {\em
  ASTRO-H}\footnote{http://astro-h.isas.jaxa.jp/en/}.  Future X-ray
missions, such as {\em
  Athena}+\footnote{\url{http://athena2.irap.omp.eu/}} and {\em
  SMART-X}\footnote{\url{http://smart-x.cfa.harvard.edu/}} will
further extend the study of the non-equilibrium phenomena into the
virialization regions of the outskirts of galaxy clusters and the
intergalactic medium.

\acknowledgments We thank Xun Shi, Eiichiro Komatsu, and the anonymous referee for useful
discussion and comments on the manuscript. This work is supported by
NSF grant AST-1412768 \& 1009811, NASA ATP grant NNX11AE07G, NASA
Chandra grants GO213004B and TM4-15007X, and by the facilities and
staff of the Yale University Faculty of Arts and Sciences High
Performance Computing Center. CA acknowledges support from the NSF
Graduate Student Research Fellowship and Alan D. Bromley Fellowship
from Yale University.

\bibliography{ms}

\begin{thebibliography}{}
\expandafter\ifx\csname natexlab\endcsname\relax\def\natexlab#1{#1}\fi

\bibitem[{{Allen} {et~al.}(2011){Allen}, {Evrard}, \& {Mantz}}]{allen_etal11}
{Allen}, S.~W., {Evrard}, A.~E., \& {Mantz}, A.~B. 2011,
  \href{http://dx.doi.org/10.1146/annurev-astro-081710-102514}{\araa},
  \href{http://adsabs.harvard.edu/abs/2011ARA%26A..49..409A}{49},
  \href{http://adsabs.harvard.edu/abs/2011ARA%26A..49..409A}{409}

\bibitem[{{Avestruz} {et~al.}(2014){Avestruz}, {Lau}, {Nagai}, \&
  {Vikhlinin}}]{avestruz_etal14}
{Avestruz}, C., {Lau}, E.~T., {Nagai}, D., \& {Vikhlinin}, A. 2014,
  \href{http://dx.doi.org/10.1088/0004-637X/791/2/117}{\apj},
  \href{http://adsabs.harvard.edu/abs/2014ApJ...791..117A}{791},
  \href{http://adsabs.harvard.edu/abs/2014ApJ...791..117A}{117}

\bibitem[{{Bautz} {et~al.}(2009){Bautz}, {Miller}, {Sanders}, {Arnaud},
  {Mushotzky}, {Porter}, {Hayashida}, {Henry}, {Hughes}, {Kawaharada},
  {Makashima}, {Sato}, \& {Tamura}}]{bautz_etal09}
{Bautz}, M.~W., {Miller}, E.~D., {Sanders}, J.~S., {et~al.} 2009, \pasj,
  \href{http://adsabs.harvard.edu/abs/2009PASJ...61.1117B}{61},
  \href{http://adsabs.harvard.edu/abs/2009PASJ...61.1117B}{1117}

\bibitem[{{Bhattacharya} {et~al.}(2012){Bhattacharya}, {Nagai}, {Shaw},
  {Crawford}, \& {Holder}}]{bhattacharya_etal2012}
{Bhattacharya}, S., {Nagai}, D., {Shaw}, L., {Crawford}, T., \& {Holder}, G.~P.
  2012, \href{http://dx.doi.org/10.1088/0004-637X/760/1/5}{\apj},
  \href{http://adsabs.harvard.edu/abs/2012ApJ...760....5B}{760},
  \href{http://adsabs.harvard.edu/abs/2012ApJ...760....5B}{5}

\bibitem[{{Bykov} {et~al.}(2008){Bykov}, {Paerels}, \&
  {Petrosian}}]{bykov_etal08}
{Bykov}, A.~M., {Paerels}, F.~B.~S., \& {Petrosian}, V. 2008,
  \href{http://dx.doi.org/10.1007/s11214-008-9309-4}{\ssr},
  \href{http://adsabs.harvard.edu/abs/2008SSRv..134..141B}{134},
  \href{http://adsabs.harvard.edu/abs/2008SSRv..134..141B}{141}

\bibitem[{{Crawford} {et~al.}(2014){Crawford}, {Schaffer}, {Bhattacharya},
  {Aird}, {Benson}, {Bleem}, {Carlstrom}, {Chang}, {Cho}, {Crites}, {de Haan},
  {Dobbs}, {Dudley}, {George}, {Halverson}, {Holder}, {Holzapfel}, {Hoover},
  {Hou}, {Hrubes}, {Keisler}, {Knox}, {Lee}, {Leitch}, {Lueker}, {Luong-Van},
  {McMahon}, {Mehl}, {Meyer}, {Millea}, {Mocanu}, {Mohr}, {Montroy}, {Padin},
  {Plagge}, {Pryke}, {Reichardt}, {Ruhl}, {Sayre}, {Shaw}, {Shirokoff},
  {Spieler}, {Staniszewski}, {Stark}, {Story}, {van Engelen}, {Vanderlinde},
  {Vieira}, {Williamson}, \& {Zahn}}]{crawford_etal2014}
{Crawford}, T.~M., {Schaffer}, K.~K., {Bhattacharya}, S., {et~al.} 2014,
  \href{http://dx.doi.org/10.1088/0004-637X/784/2/143}{\apj},
  \href{http://adsabs.harvard.edu/abs/2014ApJ...784..143C}{784},
  \href{http://adsabs.harvard.edu/abs/2014ApJ...784..143C}{143}

\bibitem[{{Diemer} \& {Kravtsov}(2014)}]{diemerandkravtsov_14}
{Diemer}, B., \& {Kravtsov}, A.~V. 2014,
  \href{http://dx.doi.org/10.1088/0004-637X/789/1/1}{\apj},
  \href{http://adsabs.harvard.edu/abs/2014ApJ...789....1D}{789},
  \href{http://adsabs.harvard.edu/abs/2014ApJ...789....1D}{1}

\bibitem[{{Ettori} \& {Fabian}(1998)}]{ettoriandfabian_98}
{Ettori}, S., \& {Fabian}, A.~C. 1998,
  \href{http://dx.doi.org/10.1046/j.1365-8711.1998.01253.x}{\mnras},
  \href{http://adsabs.harvard.edu/abs/1998MNRAS.293L..33E}{293},
  \href{http://adsabs.harvard.edu/abs/1998MNRAS.293L..33E}{L33}

\bibitem[{{Fox} \& {Loeb}(1997)}]{foxandloeb_97}
{Fox}, D.~C., \& {Loeb}, A. 1997, \apj,
  \href{http://adsabs.harvard.edu/abs/1997ApJ...491..459F}{491},
  \href{http://adsabs.harvard.edu/abs/1997ApJ...491..459F}{459}

\bibitem[{{Hill} \& {Pajer}(2013)}]{hillandpajer_13}
{Hill}, J.~C., \& {Pajer}, E. 2013,
  \href{http://dx.doi.org/10.1103/PhysRevD.88.063526}{\prd},
  \href{http://adsabs.harvard.edu/abs/2013PhRvD..88f3526H}{88},
  \href{http://adsabs.harvard.edu/abs/2013PhRvD..88f3526H}{063526}

\bibitem[{{Hill} \& {Sherwin}(2013)}]{hillandsherwin2013}
{Hill}, J.~C., \& {Sherwin}, B.~D. 2013,
  \href{http://dx.doi.org/10.1103/PhysRevD.87.023527}{\prd},
  \href{http://adsabs.harvard.edu/abs/2013PhRvD..87b3527H}{87},
  \href{http://adsabs.harvard.edu/abs/2013PhRvD..87b3527H}{023527}

\bibitem[{{Hoshino} {et~al.}(2010){Hoshino}, {Henry}, {Sato}, {Akamatsu},
  {Yokota}, {Sasaki}, {Ishisaki}, {Ohashi}, {Bautz}, {Fukazawa}, {Kawano},
  {Furuzawa}, {Hayashida}, {Tawa}, {Hughes}, {Kokubun}, \&
  {Tamura}}]{hoshino_etal10}
{Hoshino}, A., {Henry}, J.~P., {Sato}, K., {et~al.} 2010, \pasj,
  \href{http://adsabs.harvard.edu/abs/2010PASJ...62..371H}{62},
  \href{http://adsabs.harvard.edu/abs/2010PASJ...62..371H}{371}

\bibitem[{{Kaastra} \& {Jansen}(1993)}]{kaastra_etal93}
{Kaastra}, J.~S., \& {Jansen}, F.~A. 1993, \aaps,
  \href{http://adsabs.harvard.edu/abs/1993A%26AS...97..873K}{97},
  \href{http://adsabs.harvard.edu/abs/1993A%26AS...97..873K}{873}

\bibitem[{{Kawaharada} {et~al.}(2010){Kawaharada}, {Okabe}, {Umetsu},
  {Takizawa}, {Matsushita}, {Fukazawa}, {Hamana}, {Miyazaki}, {Nakazawa}, \&
  {Ohashi}}]{kawaharada_etal10}
{Kawaharada}, M., {Okabe}, N., {Umetsu}, K., {et~al.} 2010,
  \href{http://dx.doi.org/10.1088/0004-637X/714/1/423}{\apj},
  \href{http://adsabs.harvard.edu/abs/2010ApJ...714..423K}{714},
  \href{http://adsabs.harvard.edu/abs/2010ApJ...714..423K}{423}

\bibitem[{{Kravtsov}(1999)}]{kravtsov_99}
{Kravtsov}, A.~V. 1999, PhD thesis, New Mexico State Univ.

\bibitem[{{Kravtsov} {et~al.}(2002){Kravtsov}, {Klypin}, \&
  {Hoffman}}]{kravtsov_etal02}
{Kravtsov}, A.~V., {Klypin}, A., \& {Hoffman}, Y. 2002,
  \href{http://dx.doi.org/10.1086/340046}{\apj},
  \href{http://adsabs.harvard.edu/cgi-bin/nph-bib_query?bibcode=2002ApJ...571..563K}{571},
  \href{http://adsabs.harvard.edu/cgi-bin/nph-bib_query?bibcode=2002ApJ...571..563K}{563}

\bibitem[{{Lau} {et~al.}(2014){Lau}, {Nagai}, {Avestruz}, {Nelson}, \&
  {Vikhlinin}}]{lau_etal14}
{Lau}, E.~T., {Nagai}, D., {Avestruz}, C., {Nelson}, K., \& {Vikhlinin}, A.
  2014, ArXiv e-prints,
  arXiv:\href{http://adsabs.harvard.edu/abs/2014arXiv1411.5361L}{1411.5361}

\bibitem[{{Liedahl} {et~al.}(1995){Liedahl}, {Osterheld}, \&
  {Goldstein}}]{liedahl_etal95}
{Liedahl}, D.~A., {Osterheld}, A.~L., \& {Goldstein}, W.~H. 1995,
  \href{http://dx.doi.org/10.1086/187729}{\apjl},
  \href{http://adsabs.harvard.edu/abs/1995ApJ...438L.115L}{438},
  \href{http://adsabs.harvard.edu/abs/1995ApJ...438L.115L}{L115}

\bibitem[{{Mantz} {et~al.}(2010){Mantz}, {Allen}, {Rapetti}, \&
  {Ebeling}}]{mantz_etal10}
{Mantz}, A., {Allen}, S.~W., {Rapetti}, D., \& {Ebeling}, H. 2010,
  \href{http://dx.doi.org/10.1111/j.1365-2966.2010.16992.x}{\mnras},
  \href{http://adsabs.harvard.edu/abs/2010MNRAS.406.1759M}{406},
  \href{http://adsabs.harvard.edu/abs/2010MNRAS.406.1759M}{1759}

\bibitem[{{Markevitch}(2006)}]{markevitch06}
{Markevitch}, M. 2006, in ESA Special Publication, Vol. 604, The X-ray Universe
  2005, ed. A.~{Wilson}, 723,
  \href{http://adsabs.harvard.edu/abs/2006ESASP.604..723M}{astro-ph/0511345}

\bibitem[{{McDonald} {et~al.}(2014){McDonald}, {Benson}, {Vikhlinin}, {Aird},
  {Allen}, {Bautz}, {Bayliss}, {Bleem}, {Bocquet}, {Brodwin}, {Carlstrom},
  {Chang}, {Cho}, {Clocchiatti}, {Crawford}, {Crites}, {de Haan}, {Dobbs},
  {Foley}, {Forman}, {George}, {Gladders}, {Gonzalez}, {Halverson},
  {Hlavacek-Larrondo}, {Holder}, {Holzapfel}, {Hrubes}, {Jones}, {Keisler},
  {Knox}, {Lee}, {Leitch}, {Liu}, {Lueker}, {Luong-Van}, {Mantz}, {Marrone},
  {McMahon}, {Meyer}, {Miller}, {Mocanu}, {Mohr}, {Murray}, {Padin}, {Pryke},
  {Reichardt}, {Rest}, {Ruhl}, {Saliwanchik}, {Saro}, {Sayre}, {Schaffer},
  {Shirokoff}, {Spieler}, {Stalder}, {Stanford}, {Staniszewski}, {Stark},
  {Story}, {Stubbs}, {Vanderlinde}, {Vieira}, {Williamson}, {Zahn}, \&
  {Zenteno}}]{mcdonald_etal14}
{McDonald}, M., {Benson}, B.~A., {Vikhlinin}, A., {et~al.} 2014,
  \href{http://dx.doi.org/10.1088/0004-637X/794/1/67}{\apj},
  \href{http://adsabs.harvard.edu/abs/2014ApJ...794...67M}{794},
  \href{http://adsabs.harvard.edu/abs/2014ApJ...794...67M}{67}

\bibitem[{{Mewe} {et~al.}(1985){Mewe}, {Gronenschild}, \& {van den
  Oord}}]{mewe_etal85}
{Mewe}, R., {Gronenschild}, E.~H.~B.~M., \& {van den Oord}, G.~H.~J. 1985,
  \aaps, \href{http://adsabs.harvard.edu/abs/1985A%26AS...62..197M}{62},
  \href{http://adsabs.harvard.edu/abs/1985A%26AS...62..197M}{197}

\bibitem[{{Nagai} \& {Lau}(2011)}]{nagaiandlau_11}
{Nagai}, D., \& {Lau}, E.~T. 2011,
  \href{http://dx.doi.org/10.1088/2041-8205/731/1/L10}{\apjl},
  \href{http://adsabs.harvard.edu/abs/2011ApJ...731L..10N}{731},
  \href{http://adsabs.harvard.edu/abs/2011ApJ...731L..10N}{L10}

\bibitem[{{Nagai} {et~al.}(2007){Nagai}, {Vikhlinin}, \&
  {Kravtsov}}]{nagai_etal07b}
{Nagai}, D., {Vikhlinin}, A., \& {Kravtsov}, A.~V. 2007,
  \href{http://dx.doi.org/10.1086/509868}{\apj},
  \href{http://adsabs.harvard.edu/abs/2007ApJ...655...98N}{655},
  \href{http://adsabs.harvard.edu/abs/2007ApJ...655...98N}{98}

\bibitem[{{Nelson} {et~al.}(2014{\natexlab{a}}){Nelson}, {Lau}, \&
  {Nagai}}]{nelson_etal14b}
{Nelson}, K., {Lau}, E.~T., \& {Nagai}, D. 2014{\natexlab{a}},
  \href{http://dx.doi.org/10.1088/0004-637X/792/1/25}{\apj},
  \href{http://adsabs.harvard.edu/abs/2014ApJ...792...25N}{792},
  \href{http://adsabs.harvard.edu/abs/2014ApJ...792...25N}{25}

\bibitem[{{Nelson} {et~al.}(2014{\natexlab{b}}){Nelson}, {Lau}, {Nagai},
  {Rudd}, \& {Yu}}]{nelson_etal14}
{Nelson}, K., {Lau}, E.~T., {Nagai}, D., {Rudd}, D.~H., \& {Yu}, L.
  2014{\natexlab{b}},
  \href{http://dx.doi.org/10.1088/0004-637X/782/2/107}{\apj},
  \href{http://adsabs.harvard.edu/abs/2014ApJ...782..107N}{782},
  \href{http://adsabs.harvard.edu/abs/2014ApJ...782..107N}{107}

\bibitem[{{Planck Collaboration} {et~al.}(2014){Planck Collaboration}, {Ade},
  {Aghanim}, {Armitage-Caplan}, {Arnaud}, {Ashdown}, {Atrio-Barandela},
  {Aumont}, {Baccigalupi}, {Banday}, \& et~al.}]{planck_XX14}
{Planck Collaboration}, {Ade}, P.~A.~R., {Aghanim}, N., {et~al.} 2014,
  \href{http://dx.doi.org/10.1051/0004-6361/201321521}{\aap},
  \href{http://adsabs.harvard.edu/abs/2014A%26A...571A..20P}{571},
  \href{http://adsabs.harvard.edu/abs/2014A%26A...571A..20P}{A20}

\bibitem[{{Planck Collaboration Int.\ V}(2013)}]{planck_interV2013}
{Planck Collaboration Int.\ V}. 2013,
  \href{http://dx.doi.org/10.1051/0004-6361/201220040}{\aap},
  \href{http://adsabs.harvard.edu/abs/2013A%26A...550A.131P}{550},
  \href{http://adsabs.harvard.edu/abs/2013A%26A...550A.131P}{A131}

\bibitem[{{Reiprich} {et~al.}(2009){Reiprich}, {Hudson}, {Zhang}, {Sato},
  {Ishisaki}, {Hoshino}, {Ohashi}, {Ota}, \& {Fujita}}]{reiprich_etal09}
{Reiprich}, T.~H., {Hudson}, D.~S., {Zhang}, Y.-Y., {et~al.} 2009,
  \href{http://dx.doi.org/10.1051/0004-6361/200810404}{\aap},
  \href{http://adsabs.harvard.edu/abs/2009A%26A...501..899R}{501},
  \href{http://adsabs.harvard.edu/abs/2009A%26A...501..899R}{899}

\bibitem[{{Roncarelli} {et~al.}(2013){Roncarelli}, {Ettori}, {Borgani},
  {Dolag}, {Fabjan}, \& {Moscardini}}]{roncarelli_etal13}
{Roncarelli}, M., {Ettori}, S., {Borgani}, S., {et~al.} 2013,
  \href{http://dx.doi.org/10.1093/mnras/stt654}{\mnras},
  \href{http://adsabs.harvard.edu/abs/2013MNRAS.432.3030R}{432},
  \href{http://adsabs.harvard.edu/abs/2013MNRAS.432.3030R}{3030}

\bibitem[{{Rudd} \& {Nagai}(2009)}]{ruddandnagai_09}
{Rudd}, D.~H., \& {Nagai}, D. 2009,
  \href{http://dx.doi.org/10.1088/0004-637X/701/1/L16}{\apjl},
  \href{http://adsabs.harvard.edu/abs/2009ApJ...701L..16R}{701},
  \href{http://adsabs.harvard.edu/abs/2009ApJ...701L..16R}{L16}

\bibitem[{{Rudd} {et~al.}(2008){Rudd}, {Zentner}, \& {Kravtsov}}]{rudd_etal08}
{Rudd}, D.~H., {Zentner}, A.~R., \& {Kravtsov}, A.~V. 2008,
  \href{http://dx.doi.org/10.1086/523836}{\apj},
  \href{http://adsabs.harvard.edu/abs/2008ApJ...672...19R}{672},
  \href{http://adsabs.harvard.edu/abs/2008ApJ...672...19R}{19}

\bibitem[{{Shi} \& {Komatsu}(2014)}]{shiandkomatsu14}
{Shi}, X., \& {Komatsu}, E. 2014,
  \href{http://dx.doi.org/10.1093/mnras/stu858}{\mnras},
  \href{http://adsabs.harvard.edu/abs/2014MNRAS.442..521S}{442},
  \href{http://adsabs.harvard.edu/abs/2014MNRAS.442..521S}{521}

\bibitem[{{Shi} {et~al.}(2014){Shi}, {Komatsu}, {Nelson}, \&
  {Nagai}}]{shi_etal14}
{Shi}, X., {Komatsu}, E., {Nelson}, K., \& {Nagai}, D. 2014, ArXiv e-prints,
  arXiv:\href{http://adsabs.harvard.edu/abs/2014arXiv1408.3832S}{1408.3832}

\bibitem[{{Simionescu} {et~al.}(2011){Simionescu}, {Allen}, {Mantz}, {Werner},
  {Takei}, {Morris}, {Fabian}, {Sanders}, {Nulsen}, {George}, \&
  {Taylor}}]{simionescu_etal11}
{Simionescu}, A., {Allen}, S.~W., {Mantz}, A., {et~al.} 2011,
  \href{http://dx.doi.org/10.1126/science.1200331}{Science},
  \href{http://adsabs.harvard.edu/abs/2011Sci...331.1576S}{331},
  \href{http://adsabs.harvard.edu/abs/2011Sci...331.1576S}{1576}

\bibitem[{{Spitzer}(1962)}]{spitzer_62}
{Spitzer}, L. 1962, {Physics of Fully Ionized Gases} (New York: Interscience)

\bibitem[{{Tozzi} \& {Norman}(2001)}]{tozziandnorman_01}
{Tozzi}, P., \& {Norman}, C. 2001,
  \href{http://dx.doi.org/10.1086/318237}{\apj},
  \href{http://adsabs.harvard.edu/abs/2001ApJ...546...63T}{546},
  \href{http://adsabs.harvard.edu/abs/2001ApJ...546...63T}{63}

\bibitem[{{Urban} {et~al.}(2014){Urban}, {Simionescu}, {Werner}, {Allen},
  {Ehlert}, {Zhuravleva}, {Morris}, {Fabian}, {Mantz}, {Nulsen}, {Sanders}, \&
  {Takei}}]{urban_etal14}
{Urban}, O., {Simionescu}, A., {Werner}, N., {et~al.} 2014,
  \href{http://dx.doi.org/10.1093/mnras/stt2209}{\mnras},
  \href{http://adsabs.harvard.edu/abs/2014MNRAS.437.3939U}{437},
  \href{http://adsabs.harvard.edu/abs/2014MNRAS.437.3939U}{3939}

\bibitem[{{Vazza} {et~al.}(2013){Vazza}, {Eckert}, {Simionescu}, {Br{\"u}ggen},
  \& {Ettori}}]{vazza_etal13}
{Vazza}, F., {Eckert}, D., {Simionescu}, A., {Br{\"u}ggen}, M., \& {Ettori}, S.
  2013, \href{http://dx.doi.org/10.1093/mnras/sts375}{\mnras},
  \href{http://adsabs.harvard.edu/abs/2013MNRAS.429..799V}{429},
  \href{http://adsabs.harvard.edu/abs/2013MNRAS.429..799V}{799}

\bibitem[{{Vikhlinin} {et~al.}(1998){Vikhlinin}, {McNamara}, {Forman}, {Jones},
  {Quintana}, \& {Hornstrup}}]{vikhlinin_etal98}
{Vikhlinin}, A., {McNamara}, B.~R., {Forman}, W., {et~al.} 1998,
  \href{http://dx.doi.org/10.1086/305951}{\apj},
  \href{http://adsabs.harvard.edu/abs/1998ApJ...502..558V}{502},
  \href{http://adsabs.harvard.edu/abs/1998ApJ...502..558V}{558}

\bibitem[{{Vikhlinin} {et~al.}(2009){Vikhlinin}, {Kravtsov}, {Burenin},
  {Ebeling}, {Forman}, {Hornstrup}, {Jones}, {Murray}, {Nagai}, {Quintana}, \&
  {Voevodkin}}]{vikhlinin_etal09}
{Vikhlinin}, A., {Kravtsov}, A.~V., {Burenin}, R.~A., {et~al.} 2009,
  \href{http://dx.doi.org/10.1088/0004-637X/692/2/1060}{\apj},
  \href{http://adsabs.harvard.edu/abs/2009ApJ...692.1060V}{692},
  \href{http://adsabs.harvard.edu/abs/2009ApJ...692.1060V}{1060}

\bibitem[{{Voit} {et~al.}(2003){Voit}, {Balogh}, {Bower}, {Lacey}, \&
  {Bryan}}]{voit_etal03}
{Voit}, G.~M., {Balogh}, M.~L., {Bower}, R.~G., {Lacey}, C.~G., \& {Bryan},
  G.~L. 2003, \href{http://dx.doi.org/10.1086/376499}{\apj},
  \href{http://adsabs.harvard.edu/abs/2003ApJ...593..272V}{593},
  \href{http://adsabs.harvard.edu/abs/2003ApJ...593..272V}{272}

\bibitem[{{Walker} {et~al.}(2013){Walker}, {Fabian}, {Sanders}, {Simionescu},
  \& {Tawara}}]{walker_etal13}
{Walker}, S.~A., {Fabian}, A.~C., {Sanders}, J.~S., {Simionescu}, A., \&
  {Tawara}, Y. 2013, \href{http://dx.doi.org/10.1093/mnras/stt497}{\mnras},
  \href{http://adsabs.harvard.edu/abs/2013MNRAS.432..554W}{432},
  \href{http://adsabs.harvard.edu/abs/2013MNRAS.432..554W}{554}

\bibitem[{{Wong} \& {Sarazin}(2009)}]{wongandsarazin_09}
{Wong}, K.-W., \& {Sarazin}, C.~L. 2009,
  \href{http://dx.doi.org/10.1088/0004-637X/707/2/1141}{\apj},
  \href{http://adsabs.harvard.edu/abs/2009ApJ...707.1141W}{707},
  \href{http://adsabs.harvard.edu/abs/2009ApJ...707.1141W}{1141}

\bibitem[{{Zhuravleva} {et~al.}(2013){Zhuravleva}, {Churazov}, {Kravtsov},
  {Lau}, {Nagai}, \& {Sunyaev}}]{zhuravleva_etal13}
{Zhuravleva}, I., {Churazov}, E., {Kravtsov}, A., {et~al.} 2013,
  \href{http://dx.doi.org/10.1093/mnras/sts275}{\mnras},
  \href{http://adsabs.harvard.edu/abs/2013MNRAS.428.3274Z}{428},
  \href{http://adsabs.harvard.edu/abs/2013MNRAS.428.3274Z}{3274}

\end{thebibliography}

\end{document}